# Anomalous transport and thermoelectric performances of CuAgSe compounds

A. J. Hong[1], L. Li[1], H. X. Zhu[1], X. H. Zhou[1], Q. Y. He[2], W. S. Liu[3], Z. B. Yan[1], J. –M. Liu[1], and Z. F. Ren[3]

[1]*Laboratory of Solid State Microstructure, Nanjing University, Nanjing 210093, China*

[2]*Institute for Advanced Materials and Laboratory of Quantum Engineering and Quantum Materials, South China Normal University, 510006 Guangzhou, China*

[3]*Department of Physics and TcSUH, University of Houston, Houston, Texas, 77204, USA*

[**Abstract**] The copper silver selenide has two phases: the low-temperature semimetal phase ($\alpha$-CuAgSe) and high-temperature phonon-glass superionic phase ($\beta$-CuAgSe). In this work, the electric transport and thermoelectric properties of the two phases are investigated. It is revealed that the $\beta$-CuAgSe is a *p*-type semiconductor and exhibits low thermal conductivity while the $\alpha$-CuAgSe shows metallic conduction with dominant *n*-type carriers and low electrical resistivity. The thermoelectric figure of merit *zT* of the polycrystalline $\beta$-CuAgSe at 623 K is ~0.95, suggesting that superionic CuAgSe can be a promising thermoelectric candidate in the intermediate temperature range.

**I. Introduction**

The thermoelectric (TE) energy conversion technology has received much attention in the past decades.[1-9] For a TE device sandwiched between a cool bath with temperature $T_c$ and a heat bath with temperature $T_h$, the theoretical maximum conversion efficiency is governed by:[10]

$$\eta = \frac{T_h - T_c}{T_h} \frac{M-1}{M+\frac{T_c}{T_h}}, \quad M = \sqrt{1+\frac{1}{2}z(T_h+T_c)} = \sqrt{1+zT}, \quad (1)$$

where $zT$ is the dimensionless parameter known as the figure of merit, which can be related to the physical properties of TE materials:

$$zT = \frac{S^2 T}{\rho(\kappa_l + \kappa_e)}, \quad (2)$$

where $S$, $T$, $\rho$, $\kappa_l$, and $\kappa_e$ stand for thermopower (Seeback coefficient), absolute temperature, electrical resistivity, lattice and electronic thermal conductivities, respectively. For a promising TE material, a $zT$ value as high as possible is appealed via various approaches in terms of materials design and microstructure synthesis.

In the past decades, substantial efforts have been made in searching for promising TE materials of high performance including classical semiconductors such as $PbTe$, $CoSb_3$, $SiGe$ etc.[2] Along this line, some superionic semiconducting compounds have recently been demonstrated to be promising candidates for TE applications. A good example is $Cu_2Se$ compound. It is noted that $Cu_2Se$ has two phases: the low-$T$ $\alpha$-phase and high-$T$ $\beta$-phase. The low-$T$ $\alpha$-phase is very complicated in structure and diverging crystallographic data can be found in literature. Yamamoto *et al*[11] proposed a structural model for $\alpha$-$Cu_2Se$ which follows the $Fm\bar{3}m$ lattice group with fixed Cu occupation in the *fcc* lattice skeleton of the Se atoms. Fig. 1(a) shows a schematic of the lattice structure of the high-$T$ $\beta$-phase where Se atoms keep the *fcc* lattice but each Cu atom occupies randomly one of the four vertices and center of a tetrahedral unit inside the Se *fcc* unit. In this case, the Cu atoms become random in occupation (occupation-disordered) after the $\alpha$ to $\beta$ structural transformation at $T$=400 K, forming the so-called 'phonon-glass' superionic phase ($\beta$-$Cu_2Se$) in the high-$T$ range.[12-15] Both the $\alpha$-phase and $\beta$-phase have relatively low electric resistivity ($\rho$). The high-$T$ $\beta$-phase was reported to have a *zT* value as high as ~1.5 at *T* ~1000 K.[9] The underlying physics lies in that the occupation-disordered Cu atoms can easily move in the lattice under external fields (temperature field or electric field), allowing strong scattering of lattice phonons and thus low lattice thermal conductivity ($\kappa_l$), while the low electric resistivity ($\rho$) is substantially retained.

The excellent TE performances of $Cu_2Se$ compound have stimulated substantial efforts in searching for more TE materials with similar superionic structure and preferred TE performances. An immediate candidate is CuAgSe, i.e., randomly Ag-substituted $Cu_2Se$ compounds. Similar to $Cu_2Se$, CuAgSe also has two phases: the low-*T* $\alpha$-phase and high-*T* $\beta$-phase.[16-21] However, different from the two phases of $Cu_2Se$ which show similar lattice structures, the lattice structures of the two CuAgSe phases are substantially different, as schematically shown in Fig. 1(b) and (c), respectively. The low-*T* $\alpha$-phase has a unique layered structure consisting of alternating stacking of the Ag and CuSe layers, allowing probably high mobility of Ag atoms. The $\beta$-phase is similar to the superionic structure of $Cu_2Se$ with the mobile Ag/Cu cations randomly distributing among the tetrahedral sites. This may allow even lower lattice thermal conductivity than $\beta$-$Cu_2Se$, due to the random half-substitution of Cu ions by Ag ions and thus higher degree of occupation disorder, and thus an investigation of the structural and TE behaviors is interested.

Also different from the low-*T* phase of $Cu_2Se$, it is noted that the CuAgSe at low *T* is a semimetal.[17] Usually, semimetals are not preferred for high *zT* value, since the thermal broadening of the Fermi distribution function for most semimetals almost equally activates two kinds of carriers with opposite signs. This effect results in a low thermopower (*S*). However, first-principles calculations predict that CuAgSe phase possesses a significant electron-hole asymmetry in the band structure near the Fermi level,[17] which may benefit to thermopower enhancement. In addition, a semimetal would have lower resistivity than conventional TE semiconductors.

Unfortunately, due to the drastic difference in the structure and atom occupation, the two phases are expected to have very different electrical/thermo transports and TE properties, which are highly concerned from the point of views of TE applications of CuAgSe in comparison with $Cu_2Se$. These motivations certainly deserve for explorations, which will be the main purpose of this work. Nevertheless, instead of investigating the effects of substitution or microstructural controls, we pay more attention to the microscopic mechanisms for the electrical/thermo transports and TE properties associated with the phase transformations for the CuAgSe compounds.

## II. Experimental details

The sample preparation follows the standard procedure. Polycrystalline CuAgSe samples were prepared by melting high-purity powder of species Cu, Ag, and Se enclosed in a fused silica tube in vacuum. The powder mixture in stoichiometric ratios was heated to 573 K at a rate 1 K/min and maintained at this temperature for 24 h, and then slowly heated up to 973 K at a rate 2 K/min and maintained at 973 K for 24 h and then slowly cooled down to 773 K in 12 h and held there for 12 h. Then the spontaneous cooling down to room temperature yielded the as-prepared pre-samples. These pre-samples were ground into fine powder using an agate jar and plunger. The powder was subjected to direct current hot-press sintering (DCHP-2000A-04, USA) around 673 K under a pressure of 78 MPa for 6 min, producing ingots of 12.8 mm in diameter and 10 mm in height. Each ingot was cut into rectangular bars with dimensions of $3\times4\times8$ mm$^3$ for electrical transport measurements and disks with dimensions of 12.6 mm in diameter and 2.1 mm in thickness for thermal transport measurements.

The structure and crystallinity of the as-prepared samples were checked using the X-ray diffraction (XRD; Bruker Corporation) equipped with Cu-$K_\alpha$ radiation. The longitudinal resistivity and Hall resistivity as a function of magnetic field $B$ were measured using the standard four-probe method. The electric resistivity and thermopower were measured using an ULVAC ZEM-3 system (Japan) with a sealed chamber filled with a low pressure He ambient. The temperature difference across the sample was 25 K, while sufficient data were obtained by running the measurements from different initiating temperatures to guarantee the data reproducibility. The thermal diffusivity ($D$) and heat capacity ($C_P$) were measured using the laser-flash thermometer (LFA457, Netzsch). The total thermal conductivity was calculated using $\kappa = D\times C_P\times d$, where $d$ is the sample density. The differential scanning calorimetry (DSC) of these samples was carried out using a Netzsch STA449F3 device, at heating rate 10 K/min.

It should be mentioned that the synthesizing conditions for the samples are optimized and the data to be presented below can be well reproduced from sample to sample.

## III. Results and discussion

*3.1. Electrical transport*

The typical data on electrical resistivity ($\rho$) as a function of $T$ are plotted in Fig. 2(a). It is clearly seen that the $T$-dependence is evidenced with two distinct regions separating at $T = T_{\alpha\beta}$ ~470 K. In the low-$T$ region, a metallic conduction behavior is shown, consistent with earlier report,[16] while a semiconducting conduction is observed in the high-$T$ region. It is clear that the change of the electrical transport is associated with the $\alpha$-$\beta$ structural transformations which are evidenced with the DSC heat flow as a function of $T$ shown in Fig. 2(b). The sharp valley does appear at $T_{\alpha\beta}$ ~470 K.

If one compares the measured $\rho$ values inside the two phase regions with those for polycrystalline $Cu_2Se$, it is seen that the CuAgSe has an electrical resistivity comparable with that of the $Cu_2Se$. Typically, $Cu_2Se$ has the $\rho$ ~10 $\mu\Omega\cdot$m and 30 $\mu\Omega\cdot$m at $T$ ~323 K and 600 K, respectively, while CuAgSe has $\rho$ ~ 8 $\mu\Omega\cdot$m and ~58 $\mu\Omega\cdot$m at the two temperatures. Here the low $\rho$ values of the $\alpha$-CuAgSe are ascribed to the fact that the $\alpha$-CuAgSe phase is a semimetal and has high carrier mobility, to be confirmed below. The slightly higher electrical resistivity of the $\beta$-CuAgSe phase is due to the random mixing of Cu and Ag atoms, which brings additional scattering to electron transport with respect to $\beta$-$Cu_2Se$ phase, which is why the electrical resistivity of $\beta$-CuAgSe is higher than that of $\beta$-$Cu_2Se$.

*3.2. Two types of carriers*

In accompanying with the $\alpha$-$\beta$ transformations at $T_{\alpha\beta}$, one finds the substantial change of the thermopower (Seebeck coefficient $S$) as a function of $T$, as plotted in Fig. 2(c). The $S(T)$ increases rapidly from -95 $\mu$V/K at $T$ ~323 K to nearly zero at $T = T_S$ ~400 K, and then to ~130 $\mu$V/K at $T$~$T_{\alpha\beta}$. In the $\beta$-CuAgSe phase the $S(T)$ becomes nearly unchanged and the saturated $S$ value is ~190 $\mu$V/K. We are interested in the sign reversal of $S$ upon increasing $T$. This sign reversal from negative value at low $T$ to positive value at high $T$ was reported earlier,[16] with the reversal temperature $T_S$ ~400 K. This feature is not observed in $Cu_2Se$ and it reflects the existence of two types of carriers in the $\alpha$-phase, as claimed in earlier work.[16] The electrons as carriers are dominant in the low-$T$ range while the holes are dominant in the high-$T$ range. For a simple two-carriers system, the thermopower is written as $S=(\sigma_p S_p+\sigma_n S_n)/(\sigma_p+\sigma_n)$, where $\sigma_p$, $S_p$, $\sigma_n$, and $S_n$ are the $p$-type carrier electrical conductivity, $p$-type carrier thermopower, $n$-type carrier electrical conductivity, and $n$-type carrier thermopower, respectively. Since $S_p>0$ and $S_n<0$, the sign reversal of $S$ with increasing $T$ is reasonably understood. Towards the high-$T$ $\beta$-CuAgSe phase, the dominant carriers are the $p$-type and a positive $S$ is expected in this phase.

*3.3. High mobility of n-type carriers in $\alpha$-CuAgSe phase*

Before going to the TE properties of the high-$T$ $\beta$-CuAgSe phase, we come to investigate the electrical transport of the $\alpha$-CuAgSe phase. Given the existence of two types of carriers in the $\alpha$-phase, we measure the Hall resistivity to evaluate the density and mobility of the dominant $n$-type carriers at room temperature $T = 300$ K. To properly estimate the carrier concentration $n$ and mobility $\mu$, we assume that the magnetoresistance effect is dominated by the high-mobility carriers so that the conductivity tensor can be analyzed using the two-carrier model. In this case, $\mu B \ll 1$ is assumed for the low-mobility carriers, where $B$ is the magnetic field. In the standard geometry for the Hall effect, the conductivity tensors can be fitted by the following equations:[22]

$$\begin{aligned}\sigma_{xx}(B) &= en_{xx}\mu_{xx}\frac{1}{1+(\mu_{xx}B)^2}+C_{xx}\\ \sigma_{xy}(B) &= en_{xy}\mu_{xx}^2 B\left[\frac{1}{1+(\mu_{xy}B)^2}+C_{xy}\right]\end{aligned} \quad (3)$$

where $\sigma_{xx}$ ($\sigma_{xy}$), $n_{xx}$ ($n_{xy}$), $\mu_{xx}$ ($\mu_{xy}$), and $C_{xx}$ ($C_{xy}$), denote the conductivity tensors, the carrier concentration components, the carrier mobility components, and the low-mobility components for the conductivity tensor, respectively. The conductivity tensors are related to $\rho_{xx}$ and $\rho_{xy}$:

$$\begin{aligned}\sigma_{xx} &= \rho_{xx}/\left(\rho_{xx}^2+\rho_{xy}^2\right)\\ \sigma_{xy} &= -\rho_{xy}/\left(\rho_{xx}^2+\rho_{xy}^2\right)\end{aligned} \quad (4)$$

where $\rho_{xx}$ and $\rho_{xy}$ are the longitudinal resistivity and Hall resistivity, respectively.

The $B$-dependences of $\rho_{xx}$ and $\rho_{xy}$ at room temperature are plotted in Fig. 3(a) and (b),

respectively. Both the $\rho_{xx}$ and $\rho_{xy}$ increase with increasing $B$, as well known. These data are fitted using Eqs. (3) and (4) and the fitted data for $\sigma_{xx}$ and $\sigma_{xy}$ are presented in Fig. 3(c) and (d), respectively, where the solid lines represent the fitted curves. The as-evaluated carrier concentrations $n_{xx} = 3.9587 \times 10^{18}$ cm$^{-3}$ ($n_{xy} = 3.0996 \times 10^{18}$ cm$^{-3}$) and carrier mobility $\mu_{xx} = 2018$ cm$^2$/V·s ($\mu_{xy} = 2204$ cm$^2$/V·s). It is seen that the Hall mobility of low-$T$ $\alpha$-phase is far higher than that of $\alpha$-Cu$_2$Se ($\mu_{xx} \sim 20$ cm$^2$/V·s) while the carrier density of the low-$T$ $\alpha$-phase is much lower than that of $\alpha$-Cu$_2$Se ($n_{xx} = 3.2 \times 10^{20}$ cm$^{-3}$). Surely, this high carrier mobility has its structural origin, noting that the $\alpha$-CuAgSe phase has a layer-like structure, as shown in Fig. 1(b), consistent with the semi-metallic band structure. There is the experimental evidence for the semi-metallic band structure with the light-mass electron carriers which can result in high carrier mobility.

*3.4. Thermal transport*

We then turn to the thermal transport behavior which is critical for the TE performances. The measured thermal conductivity $\kappa$ as a function of $T$ is plotted in Fig. 4, noting that this data set is consistent from different runs in different institutes. Basically, the thermal conductivity is high in the $\alpha$-phase region and low in the $\beta$-phase region, with a gradual transition across the $\alpha$-$\beta$ phase transformations around $T_{\alpha\beta}$. Regarding the $T$-dependences of $\kappa$ in the two regions, one sees a gradual but weak increasing of $\kappa$ with increasing $T$ in the $\alpha$-phase region while no remarkable $T$-dependence in the $\beta$-phase region is observed. In particular, the $\kappa$ value of the $\beta$-phase is as low as ~0.47 W m$^{-1}$ K$^{-1}$, much lower than that of the $\beta$-Cu$_2$Se in this $T$-range (0.9 W m$^{-1}$ K$^{-1}$ at $T$ ~600 K). Such a low $\kappa$ value is highly preferable for thermoelectric applications.

A reasonable explanation for the low $\kappa$ value is the specific superionic (liquid-like) lattice structure of the $\beta$-phase in which the mixed Cu and Ag atoms distributed randomly over tetrahedral sites, which substantially weakens the thermal transport. For the low-$T$ $\alpha$-phase, the layer lattice structure however allows relatively easy phonon and electron thermal transport. Furthermore, we look at the lattice phonon and carrier contributions to the thermal conductivity. One estimates the carrier thermal conductivity $\kappa_e$ from the Wiedeman-Franz law $\kappa_e = L\sigma T$ where the Lorenz number $L = 2.1 \times 10^{-8}$ V$^{-2}$ K$^{-2}$ is taken. The evaluated lattice thermal conductivity $\kappa_l = \kappa - \kappa_e$ as a function of $T$ is plotted in Fig. 4. Several features regarding the thermal conductivity mechanism can be seen. First, in the low-$T$ $\alpha$-phase, the high $\kappa_e$ values are expected to be associated with the semimetal behavior, while the $\kappa_e$ of the high-$T$ $\beta$-phase is much lower. Second, in the high-$T$ $\beta$-phase, the $\kappa_l$ can be as low as ~0.27 W m$^{-1}$ K$^{-1}$ which are much lower than the values of $\beta$-Cu$_2$Se (0.4-0.6 W m$^{-1}$ K$^{-1}$). Local atomic jumps and rearrangement of the 'liquid' $\beta$-CuAgSe inhibit the propagation of transverse waves, thereby leading to a significant reduction in the lattice thermal conductivity.[10]

*3.5. Thermoelectric performance*

Finally, we present the figure-of-merit $zT$ as a function of $T$, as shown in Fig. 5. The $zT$ value at $T$ = 323K is ~0.28. In between 323 K and 400 K ($<T_{\alpha\beta}$), the very low $zT$ value is attributed to the existence of two types of carriers which lead to the sign reversal of thermopower $S$. However, in the $\beta$-phase region, the $zT$ value increases with increasing $T$ and reaches 0.95 at $T$ = 623 K, while the $zT$ value of Cu$_2$Se is ~0.6 at 650 K.

In addition, the phase stability of this CuAgSe compound was checked from two aspects. First, the same sample was submitted to consecutive four TE measuring runs using the ULVAC ZEM-3 system (Japan) with a sealed chamber filled with a low pressure He ambient. The measured electrical resistivity and thermopower for three runs as a function of $T$ from 300 K to 650 K are plotted in Fig. 6(a) and (b), respectively, for comparisons. No remarkable differences among these data are seen. Second, The XRD checking of this sample in between two TE measuring runs was performed, and there was no any difference of the crystalline structure observed. These data suggest that this CuAgSe compound is structurally stable up to 650 K in the present sense of characterization.

**IV. Conclusion**

In conclusion, we have investigated the structural, electrical, and thermal transport properties of CuAgSe compound. The $\alpha$-$\beta$ structural transformation happens at $T$ ~470 K, accompanied with remarkable difference in the electrical and thermal transports. It has been revealed that the low-$T$ $\alpha$-phase accommodates dominant $n$-type carriers with minor $p$-type carriers, while the high-$T$ $\beta$-phase is dominated with $p$-type carriers. The high-$T$ $\beta$-phase exhibits good thermoelectric properties, in particular very low thermal conductivity, which allow the superionic CuAgSe to be a promising candidate for potentially intermediate temperature thermoelectric applications.

**Acknowledgement:**

This work was supported by the National 973 Projects of China (Grant No. 2011CB922101), the Natural Science Foundation of China (Grants Nos. 11234005, 11374147, and 51372092), and the Priority Academic Program Development of Jiangsu Higher Education Institutions, China.

**Figure Captions**

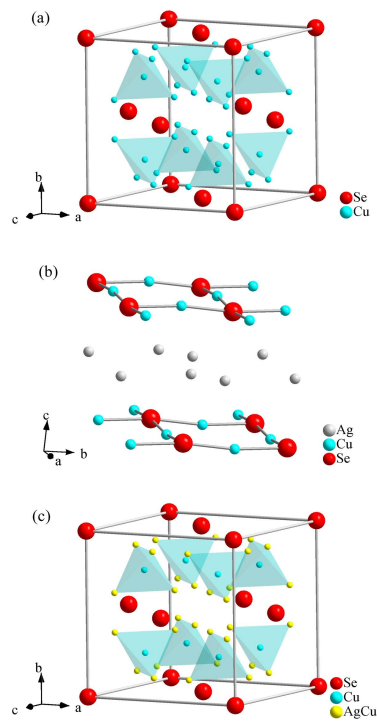

Fig. 1. (color online) Schematic drawing of lattice structures of compounds $\beta$-$Cu_2Se$ (a), $\alpha$-CuAgSe (b), and $\beta$-CuAgSe (c).

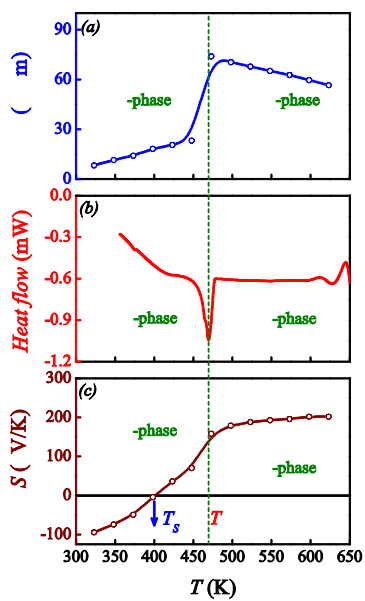

Fig. 2. (color online) Measured $T$-dependences of electrical resistivity $\rho$ (a), heat flow (b), and thermopower $S$ (c). Here, the heat flow is defined as the sample absorbed or released thermal power at a given temperature during the sample heating at a constant rate. The negative heat

flow indicates the thermal power released from the sample.

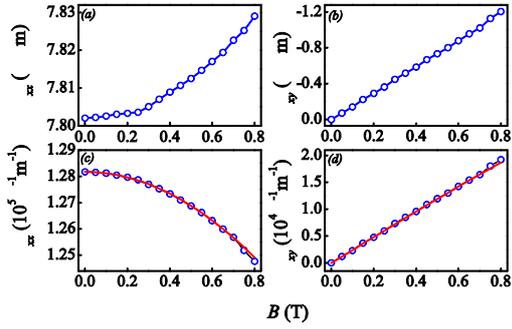

Fig. 3. (color online) Measured longitudinal electrical resistivity $\rho_{xx}$ (a) and Hall resistivity $\rho_{xy}$ (b) as a function of magnetic field $B$ at room temperature. The evaluated longitudinal conductivity $\sigma_{xx}$ (c) and Hall conductivity $\sigma_{xy}$ (d) as a function of $B$. The solid curves are the fitting data using Eqs. (3) and (4).

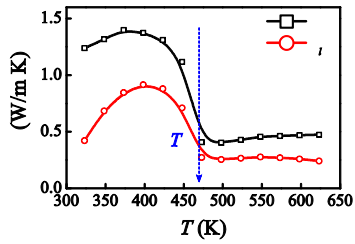

Fig. 4. (color online) Measured $T$-dependence of the thermal conductivity and extracted lattice thermal conductivity.

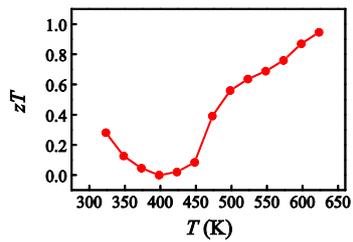

Fig. 5. (color online) Thermoelectric figure of merit $zT$ as a function of $T$.

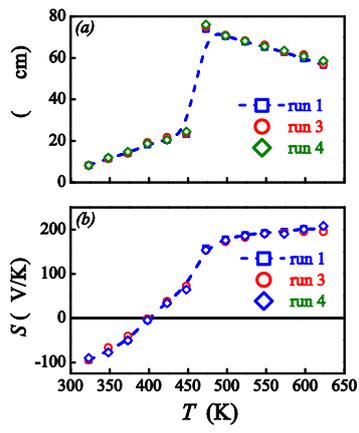

Fig. 6. (color online) Electrical resistivity $\rho$ (a) and thermopower $S$ (b) as a function of $T$, as evaluated from consecutive runs (run No. 1, 3, & 4) of the TE measurements from 300 K to 650 K.